\documentclass[prd,preprint,
superscriptaddress,tightenlines,nofootinbib, eqsecnum]{revtex4-2}

\usepackage{amsmath}
\usepackage{amsfonts}
\usepackage{amssymb}
\usepackage{bm}
\usepackage{hyperref}
\usepackage{mathrsfs}
\usepackage{graphicx}

\usepackage{empheq}

\usepackage{ulem}
\usepackage[usenames]{color}

\allowdisplaybreaks

\hypersetup{
 bookmarks=true,         % show bookmarks bar?
 unicode=false,          % non-Latin characters in Acrobat???s bookmarks
 pdftoolbar=true,        % show Acrobat???s toolbar?
 pdfmenubar=true,        % show Acrobat???s menu?
 pdffitwindow=false,     % window fit to page when opened
 pdfstartview={FitH},    % fits the width of the page to the window
 pdftitle={My title},    % title
 pdfauthor={Author},     % author
 pdfsubject={Subject},   % subject of the document
 pdfcreator={Creator},   % creator of the document
 pdfproducer={Producer}, % producer of the document
 pdfkeywords={keyword1} {key2} {key3}, % list of keywords
 pdfnewwindow=true,      % links in new window
 colorlinks=true,       % false: boxed links; true: colored links
 linkcolor=red,          % color of internal links
 citecolor=cyan,        % color of links to bibliography
 filecolor=magenta,      % color of file links
 urlcolor=darkgreen,           % color of external links
 linktocpage=true
}

\definecolor{darkgreen}{rgb}{0,0.5,0}

\newcommand{\FP}{\mathop{\mathrm{FP}}_{B=0}}
\newcommand{\FPprop}{\mathop{\mathrm{FP}}_{B=0}\Box^{-1}_\text{ret}}

\DeclareSymbolFontAlphabet{\mathrsfs}{rsfs}
\DeclareMathAlphabet{\mathcal}{OMS}{cmsy}{m}{n}

\newcommand{\dd}{\mathrm{d}}

\newcommand{\dM}{\mathrm{M}}
\newcommand{\dMbar}{\overline{\mathrm{M}}}
\newcommand{\dS}{\mathrm{S}}
\newcommand{\dSbar}{\overline{\mathrm{S}}}

\allowdisplaybreaks
\DeclareSymbolFontAlphabet{\mathrsfs}{rsfs}
\DeclareMathAlphabet{\mathcal}{OMS}{cmsy}{m}{n}

\begin{document}
	
\title{The Quadrupole Moment of Compact Binaries \\to the Fourth post-Newtonian Order:\\ Relating the Harmonic and Radiative Metrics}

\author{David \textsc{Trestini}}\email{david.trestini@obspm.fr}
\affiliation{Laboratoire Univers et Théories, Observatoire de Paris, Université PSL, Université Paris Cité, CNRS, F-92190 Meudon, France}
\affiliation{$\mathcal{G}\mathbb{R}\varepsilon{\mathbb{C}}\mathcal{O}$, 
	Institut d'Astrophysique de Paris,\\ UMR 7095, CNRS, Sorbonne Universit{\'e},\\
	98\textsuperscript{bis} boulevard Arago, 75014 Paris, France}
	
\author{Fran\c{c}ois \textsc{Larrouturou}}\email{francois.larrouturou@desy.de}
\affiliation{Deutsches Elektronen-Synchrotron DESY, Notkestr. 85, 22607 Hamburg, Germany}

\author{Luc \textsc{Blanchet}}\email{luc.blanchet@iap.fr}
\affiliation{$\mathcal{G}\mathbb{R}\varepsilon{\mathbb{C}}\mathcal{O}$, 
	Institut d'Astrophysique de Paris,\\ UMR 7095, CNRS, Sorbonne Universit{\'e},\\
	98\textsuperscript{bis} boulevard Arago, 75014 Paris, France}
\affiliation{Institut de Physique Th\'eorique, Universit\'e Paris-Saclay,\\
CEA, CNRS, 91191 Gif-sur-Yvette, France}

\date{\today}

\preprint{DESY-22-116}

\begin{abstract}
Motivated by the completion of the fourth post-Newtonian (4PN) gravitational-wave generation from compact binary systems, we analyze and contrast different constructions of the metric outside an isolated system, using post-Minkowskian expansions. The metric in ``harmonic'' coordinates has been investigated previously, in particular to compute tails and memory effects. However, it is plagued by powers of the logarithm of the radial distance $r$ when $r\to\infty$ (with $t-r/c=$ const). As a result, the tedious computation of the ``tail-of-memory'' effect, which enters the gravitational-wave flux at 4PN order, is more efficiently performed in the so-called ``radiative'' coordinates, which admit a (Bondi-type) expansion at infinity in simple powers of $r^{-1}$, without any logarithms. Here we consider a particular construction, performed order by order in the post-Minkowskian expansion, which directly yields a metric in radiative coordinates. We relate both constructions, and prove that they are physically equivalent as soon as a relation between the ``canonical'' moments which parametrize the radiative metric, and those parametrizing the harmonic metric, is verified. We provide the appropriate relation for the mass quadrupole moment at 4PN order, which will be crucial when deriving  the ``tail-of-memory'' contribution to the gravitational flux. 
\end{abstract}


\maketitle

\section{Introduction}
\label{sec:motivations}

Improving the accuracy of gravitational wave (GW) templates is one of the keys to an era of precision gravitational astronomy. Indeed, such waveforms are critical for the parameter estimation of ground-based detectors and, in the future, spaceborne ones. Among the different techniques that were developed for this purpose, the post-Newtonian (PN) approximation describes the inspiralling phase of compact binaries, and constitutes the basis for effective phenomenological methods such as EOB (effective-one-body) or IMR (inspiral-merger-ringdown), which connect the inspiral and merger phases (see~\cite{Maggiore,BlanchetLR,BuonSathya15,Porto16}). 

Crucial to derive high precision PN waveforms is the knowledge of the mass-type quadrupole moment at a high level of accuracy. This quantity has been computed, in the case of non-spinning compact binaries, at the increasingly high 1PN~\cite{WagW76,BS89}, 2PN~\cite{BDI95,BDIWW95,WW96,LMRY19} and 3PN~\cite{BIJ02,BI04mult,BDEI04,BDEI05dr} orders. At the 4PN order, the ``canonical'' mass-type quadrupole moment has been computed, properly regularized and renormalized, and finally linked to the source components~\cite{MHLMFB20,MQ4PN_IR,MQ4PN_renorm,MQ4PN_jauge}. Similarly, the mass octupole and current quadrupole have been computed up to 3PN order~\cite{FBI15,HFB_courant}.

But the quantities entering the observables, such as the GW phase, are not the ``canonical'' moments (expressed in the vicinity of the source), but the ``radiative'' ones, expressed at future null infinity $\mathcal{I}^{+}$, \emph{i.e.} $r\to\infty$, with the retarded, null or asymptotically null coordinate $u=t-r/c$ held constant. Radiative moments differ from canonical moments due to the non-linear effects that occur during the propagation of GWs, for example the \emph{tail} effect entering at 1.5PN order, due to the scattering of GWs onto the static curvature generated by the source, the non-linear \emph{memory} at 2.5PN order, due to reemission of GWs \textit{via} the scattering of GWs by the linear quadrupolar wave itself, or even the \emph{tail-of-tail} at 3PN order, due to a double scattering of the GW onto the static curvature generated by the source.

It turns out that, in order to derive the 4PN mass-type quadrupole moment, one must also perform the tedious computation of the \emph{tail-of-memory} effect (due to mass-quadrupole-quadrupole interactions), along with an easier \emph{spin-quadrupole tail} effect (due to mass-spin-quadrupole interactions), which both arise for the first time at 4PN order~\cite{TB23_ToM}. Using the multipolar-post-Minkowskian  (MPM) expansion scheme~\cite{BD86}, the computation of the radiative moment is usually performed in harmonic (or de Donder) coordinates. However harmonic coordinates are plagued by powers of the logarithm of the radial distance when $r\to\infty$ (with $t-r/c$ held constant), and extracting the quantities of interest in such a coordinate system for highly non-linear effects such as the \emph{tail-of-memory} becomes quite difficult, due to the polylogarithmic behaviour of the metric at $\mathcal{I}^{+}$.

On the other hand, it is known that GWs generated by isolated systems can be also described using ``radiative'' type coordinates, in which the metric admits an expansion at~$\mathcal{I}^{+}$ without the polylogarithmic behaviour of harmonic coordinates. In such coordinates the field equations may be formally integrated order by order when $r\to\infty$. The paragon of radiative coordinates is the Bondi coordinate system~\cite{BBM62, Sachs62} or its variant the Newman-Unti coordinate system~\cite{newman1963class}.\footnote{A detailed construction of Bondi-NU coordinates starting from harmonic coordinates can be found in~\cite{BCFOS21}.} However the class of radiative coordinate systems (either spherical or Cartesian) is very large~\cite{Papa69, MadoreI, MadoreII}.

An explicit construction of radiative coordinates using the MPM expansion was proposed in~\cite{B87}. This construction is restricted to a metric which is stationarity in the remote past, before some given instant $-\mathcal{T}$. Under this hypothesis it was proven, up to any order in a perturbative post-Minkowskian sense, that the metric admits a Bondi-type expansion at $\mathcal{I}^{+}$ to all orders in $1/r$, and that it obeys standard definitions for asymptotic flatness/simplicity at future null infinity~\cite{Penrose63,Penrose65,GH78}. 

However, the radiative MPM construction of~\cite{B87} has never been used for practical computations, until it was recently found to be extremely useful and important when implementing the difficult computation of the tails-of-memory and spin-quadrupole tails at 4PN order~\cite{TB23_ToM}.

Nevertheless, previous computations of tails, memory, tails-of-tails and even tails-of-tails-of-tails in the GW flux were done using the harmonic coordinate MPM algorithm~\cite{BD92, B98quad, B98tail, MBF16}. The aim of the present work is thus to analyze and relate the different MPM constructions of the metric in harmonic and radiative coordinates. Closely following the recent work~\cite{MQ4PN_jauge}, we prove that the relation between both descriptions boils down to a coordinate transformation and a simple redefinition of the moments parametrizing the two metrics.

As a result, we find that, in the center-of-mass frame, the mass-type ``canonical'' quadrupole moment in the radiative gauge, denoted below $\dMbar_{ij}$, is linked to the corresponding one in the harmonic gauge, say $\dM_{ij}$, by the 4PN-accurate relation presented in Eqs.~\eqref{eq:deltaMijbarres} and~\eqref{eq:deltaMijres} below, which constitute the main result of this paper.

The plan of the paper is as follows. In Sec.~\ref{sec:HarmandRad}, we review both the harmonic and radiative post-Minkowskian constructions of the metric. Sec.~\ref{sec:General} presents the general formalism to link these two constructions, by essentially adapting the results of Ref.~\cite{MQ4PN_jauge}. We review the consistency of our method in Sec.~\ref{sec:Tails} by applying it at the quadratic order to the case of the \emph{tail} interaction, and then to cubic order for the \emph{tail-of-tail} interaction. Finally, the full result at cubic order (comprising the \emph{tail-of-memory} and \emph{spin-quadrupole tail} interactions) is given in Sec.~\ref{sec:cubicRes}. Useful integration formulae complementing those in~\cite{MQ4PN_jauge} are presented in App.~\ref{sec:Retard}.

\section{The Harmonic and Radiative constructions}
\label{sec:HarmandRad}

In this paper, we consider two constructions of the general solution of the Einstein vacuum field equations outside a matter source, in the form of a MPM expansion~\cite{BD86, B87}. The vacuum field equations in a generic coordinate system are written as
\begin{equation}\label{eq:EFE_vac}
	\Box h^{\mu\nu} - \partial H^{\mu\nu} = \Lambda^{\mu\nu}\,,
\end{equation}
where the ``gothic metric'' deviation from the Minkowski metric is $h^{\mu\nu} \equiv \sqrt{-g}\,g^{\mu\nu}-\eta^{\mu\nu}$; $g^{\mu\nu}$ is the inverse of the usual covariant metric; $\eta^{\mu\nu}$, that of the Minkowski metric; $g\equiv\text{det}(g_{\mu\nu})$~is the determinant (our signature is $+2$); $\Box \equiv \Box_\eta$ is the flat d'Alembertian operator; and the gravitational source term $\Lambda^{\mu\nu}\equiv\Lambda^{\mu\nu}[h,\partial h,\partial^2h]$ is at least quadratic in $h$ and its first and second partial derivatives; finally, we pose as shorthands $H^\mu \equiv \partial_\nu h^{\mu\nu}$ and \mbox{$\partial H^{\mu\nu}\equiv \partial^\mu H^\nu + \partial^\nu H^\mu - \eta^{\mu\nu}\partial_\rho H^{\rho}$}.

In the usual harmonic gauge, we have $H^\mu = 0$ by definition, so the vacuum equations~\eqref{eq:EFE_vac} reduce to ordinary wave equations with a quadratic source term. But the purpose of this work is precisely to depart from the harmonic gauge, and thus the second term in the left-hand side of Eq.~\eqref{eq:EFE_vac} will play an important role. Since we are not restricting to harmonic coordinates, the source term also differs from the one usually given in harmonic coordinates. More specifically, the full source term here can be decomposed as
\begin{equation}
\label{LambdaSource}
\Lambda^{\mu\nu}[h] = \Lambda^{\mu\nu}_{\mathrm{harm}}[h] + 2 h^{\rho(\mu}\partial_{\rho} H^{\nu)} - \partial_\rho (h^{\mu\nu} H^{\rho})\,,
\end{equation}
where $ \Lambda^{\mu\nu}_{\mathrm{harm}}[h] $ is the source term  when assuming the harmonic gauge condition, given for example by Eq.~(24) in~\cite{BlanchetLR}. We have $\partial_\nu\partial H^{\mu\nu}=\Box H^\mu$ hence the source term of the Einstein equation~\eqref{eq:EFE_vac} is identically divergenceless, $\partial_\nu\Lambda^{\mu\nu}\equiv 0$, which is equivalent to the contracted Bianchi identity.

\subsection{The harmonic algorithm}
\label{sec:harmalg}

To begin with, we work in the harmonic gauge, hence $\partial_\nu h_\text{harm}^{\mu\nu}=0$. The MPM expansion is defined up to any post-Minkowskian (PM) order $n$ by:
\begin{equation}\label{eq:hharm}
h_\text{harm}^{\mu\nu} = \sum_{n=1}^{+\infty} G^n h^{\mu\nu}_{\text{harm}\, n}
\,.
\end{equation}
The first step to construct such solution is naturally the linearized approximation $n=1$, defined by means of two sets of symmetric-trace-free (STF) multipole moments $\{\dM_L,\dS_L\}$ (where $L=i_1 i_2\cdots i_\ell$ is the multi-index made of $\ell$ spatial indices) as~\cite{SB58, Pi64, Th80, BD86}
\begin{align}\label{eq:h1harm}
	h_{\text{harm}\,1}^{00} &= - \frac{4}{c^2}\sum_{\ell \geqslant 0}\frac{(-)^\ell}{\ell !}\partial_L\left[\frac{1}{r}\,\dM_L\left(t-\frac{r}{c}\right)\right]\,,\nonumber\\
	%%%%%%%%%%%%%%%%%%%%%%%%%%%%%%%%%%%%%%%%%%%%%%%%%%%%%%%%%
	h_{\text{harm}\,1}^{0i} &= \frac{4}{c^3}\sum_{\ell \geqslant 1}\frac{(-)^\ell}{\ell !}\left\lbrace\partial_{L-1}\left[\frac{1}{r}\,\dM_{iL-1}^{(1)}\left(t-\frac{r}{c}\right)\right]+ \frac{\ell}{\ell+1}\partial_L\left[\frac{1}{r}\,\dS_{i\vert L}\left(t-\frac{r}{c}\right)\right]\right\rbrace\,,\\
	%%%%%%%%%%%%%%%%%%%%%%%%%%%%%%%%%%%%%%%%%%%%%%%%%%%%%%%%%
	h_{\text{harm}\,1}^{ij} &= -\frac{4}{c^4}\sum_{\ell \geqslant 2}\frac{(-)^\ell}{\ell !}\left\lbrace\partial_{L-2}\left[\frac{1}{r}\,\dM_{ijL-2}^{(2)}\left(t-\frac{r}{c}\right)\right]+ \frac{2\ell}{\ell+1}\partial_{L-1}\left[\frac{1}{r}\,\dS_{(i\vert j)L-1}^{(1)}\left(t-\frac{r}{c}\right)\right]\right\rbrace\,.\nonumber
\end{align}
We have adopted the convention of~\cite{HFB_courant} for the current-type moment, which we define as $\dS_{i\vert L} \equiv \varepsilon_{ii_\ell k}\,\dS_{kL-1}$, and the horizontal bar means that the indices are excluded from the symmetrization. The STF multipole moments $\{\dM_L,\dS_L\}$ are arbitrary functions of the retarded time $t-r/c$ of the harmonic coordinates, the only constraint being that the monopole and dipoles satisfy the usual conservation laws, \emph{i.e.} that the mass monopole $\dM$, the time derivative of the mass dipole $\dM^{(1)}_i$ and the current dipole $\dS_i$ are all time-independent constants. Thus~$\dM$,~$\dM_i$~and~$\dS_i$~represent the Arnowitt-Deser-Misner (ADM) quantities, made of matter and GW~contributions. In this paper, we restrict ourselves to a center-of-mass frame defined by $\dM_i=0$.

The harmonic algorithm is defined by the ``canonical'' construction exposed in Ref.~\cite{MQ4PN_jauge}. Namely, suppose that we have constructed the $n-1$ first PM coefficients $h^{\mu\nu}_{\text{harm}\, m}$ for any $m\leqslant n-1$, all satisfying the harmonic gauge condition $\partial_\nu h^{\mu\nu}_{\text{harm}\, m}=0$. Then we construct the $n$-th order coefficient as follows. It satisfies $\Box h^{\mu\nu}_{\text{harm}\, n}=\Lambda_{\text{harm}\, n}^{\mu\nu}$ together with $\partial_\nu h^{\mu\nu}_{\text{harm}\, n}=0$, where the source term is constructed out of the previous iterations: \mbox{$\Lambda_{\text{harm}\, n}^{\mu\nu}\equiv\Lambda^{\mu\nu}[h_{\text{harm}\, m}; m\leqslant n-1]$}. We first construct a particular retarded solution of the wave equation as
\begin{align}\label{eq:unharm}
	u_{\text{harm}\, n}^{\mu\nu} 
	\equiv 
	\FPprop\biggl[\left(\frac{r}{r_0}\right)^B\Lambda_{\text{harm}\, n}^{\mu\nu} \biggr]\,.
\end{align}
Here $\Box^{-1}_\text{ret}$ is the usual retarded inverse d'Alembertian operator, and the symbol $\FP$ refers to the finite part (FP) or zeroth power coefficient in the Laurent expansion when the complex parameter $B$ tends to zero. The regularization factor $(r/r_0)^B$ multiplying the source term is a protection against the divergence of the multipole expansion when $r\to 0$. The constant length scale $r_0$ is arbitrary and has to disappear from any physical result in the end. We let the curious reader refer to~\cite{MQ4PN_jauge} for further details. Since $\partial_\nu \Lambda_{\text{harm}\, n}^{\mu\nu} = 0$, the divergence of the particular solution $u_{\text{harm}\, n}^{\mu\nu}$ reads 
\begin{equation}\label{eq:wnharm}
	w_{\text{harm}\, n}^{\mu} 
	\equiv \partial_\nu u_{\text{harm}\, n}^{\mu\nu} = 
	\FPprop\biggl[B\left(\frac{r}{r_0}\right)^B\frac{n_i}{r}\,\Lambda_{\text{harm}\, n}^{\mu i} \biggr]\,.
\end{equation}
The factor $B$ comes from the differentiation of the regulator $(r/r_0)^B$. Because of it, the term~\eqref{eq:wnharm} is non-zero only when the integral develops a pole $\propto 1/B$ when $B\to 0$. Furthermore, the coefficient of the pole is necessarily a homogeneous retarded solution of the wave equation, $\Box w_{\text{harm}\, n}^{\mu} = 0$. At this stage, we apply the MPM ``harmonicity'' algorithm to construct from $w_{\text{harm}\, n}^{\mu}$ another homogeneous retarded solution, say 
\begin{equation}\label{eq:vnharm}
	v_{\text{harm}\, n}^{\mu\nu} \equiv \mathcal{V}^{\mu\nu}\bigl[w_{\text{harm}\, n}\bigr]\,,
\end{equation}
satisfying at once $\partial_\mu v_{\text{harm}\, n}^{\mu\nu} = - w_{\text{harm}\, n}^{\mu}$ and $\Box v_{\text{harm}\, n}^{\mu\nu} = 0$. The above harmonicity algorithm $w^\mu\longrightarrow\mathcal{V}^{\mu\nu}[w]$ is explicitly defined by Eqs.~(2.11)-(2.12) in Ref.~\cite{B98quad}. Finally, the harmonic metric at order $n$, now satisfying the full Einstein vacuum equations in harmonic coordinates at the $n$-th order, reads
\begin{equation}\label{eq:hnharm}
	h_{\text{harm}\, n}^{\mu\nu} = u_{\text{harm}\, n}^{\mu\nu} + v_{\text{harm}\, n}^{\mu\nu}\,.
\end{equation}
The harmonic metric is a non-linear functional of the moments $\{\dM_L,\dS_L\}$ and represents the most general solution of the Einstein field equations in the vacuum region outside an isolated system~\cite{BD86}.

\subsection{The radiative algorithm}
\label{sec:radalg}

We now describe a different MPM algorithm, proposed in~\cite{B87}, which directly builds the metric in a radiative coordinate system,
\begin{equation}\label{eq:hrad}
	h_\text{rad}^{\mu\nu} = \sum_{n=1}^{+\infty} G^n h^{\mu\nu}_{\text{rad}\, n}
	\,.
\end{equation}
By radiative coordinate system we mean a coordinate system whose retarded time coordinate, say $u\equiv t-r/c$, is a null coordinate, \textit{i.e.} satisfies $g_\text{rad}^{\mu\nu}\partial_\mu u \partial_\nu u = 0$, or at least, becomes a null coordinate in the asymptotic limit $r\to\infty$ with $u$ held constant, \textit{i.e.} in a neighbourhood of $\mathcal{I}^{+}$~\cite{Papa69, MadoreI, MadoreII}. In such class of coordinate systems, the metric admits a Bondi-like expansion at infinity, in simple powers of the inverse distance $1/r$, without any logarithms of $r$ as would occur in harmonic coordinates.\footnote{In this paper, since we are iteratively constructing a coordinate system order by order, it is convenient to consider the coordinates as dummy variables and denote them by $(t, r)$. Even at the end, once we have obtained the full radiative metric, we shall continue to denote the radiative coordinates by the generic $(t, r)$, although it might be more appropriate to denote them by $(T, R)$, for instance.}

Even at the linearized level, it is necessary to correct the harmonic coordinate metric in order to satisfy the requirement of asymptotically null retarded time. Consequently, the radiative MPM algorithm starts by performing a linear gauge transformation of the harmonic-coordinate metric defined by~\eqref{eq:h1harm}. A crucial point is that the multipole moments that parametrize the radiative algorithm will differ from their counterparts in the harmonic algorithm. In other words, they will have a different expression when expressed explicitly in terms of the source \emph{via} a matched asymptotic expansion procedure. This observation leads us to define the radiative algorithm using different multipole moments, which we will note~$\{\dMbar_L,\overline{\dS}_L\}$. Thus, at linear order we pose
\begin{equation}\label{eq:hrad1}
	h^{\mu\nu}_{\text{rad}\, 1} = h^{\mu\nu}_{\text{harm}\, 1}\bigl[\dMbar_L,\overline{\dS}_L\bigr]+ \partial \xi_1^{\mu\nu}
	\,,
\end{equation}
where $h^{\mu\nu}_{\text{harm}\, 1}\bigl[\dMbar_L,\overline{\dS}_L\bigr]$ has exactly the same functional expression as in harmonic coordinates, given by Eqs.~\eqref{eq:h1harm}, but is now computed with the set of moments $\{\dMbar_L,\overline{\dS}_L\}$. The linear gauge transformation $\partial \xi_1^{\mu\nu}\equiv \partial^\mu \xi_1^\nu + \partial^\nu \xi_1^\mu - \eta^{\mu\nu}\partial_\rho \xi_1^{\rho}$ is defined by the gauge vector
\begin{equation}\label{eq:xi1}
	\xi^\mu_{1} = \frac{2 \dM}{c^2} \, \eta^{0\mu}\ln
	\left(\frac{r}{b_0}\right) \,,
	%= \left( - \frac{2 \dM}{c^2} \, \ln	\left(\frac{r}{r_0}\right), \bm{0}\right)\,.
\end{equation}
where $\dM$ is the mass monopole associated with the set of moments, $b_0$ denotes an arbitrary length scale, and we have $\eta^{0\mu}=(-1,\bm{0})$ with our signature. Since the gauge vector will only appear in the derivative form $\partial \xi_1^{\mu\nu}$ in the radiative algorithm, the unphysical scale $b_0$ will actually never enter the radiative metric. However we shall prove that $b_0$ is identical to the scale which is used in harmonic constructions of the metric when building the observable quantities at infinity, \emph{via} the radiative multipole moments, see \emph{e.g.}~\cite{FBI15}.  

Note that although the two sets of multipole moments $\{\dM_L,\dS_L\}$ and $\{\dMbar_L,\overline{\dS}_L\}$ will differ in general (as we shall compute explicitly below), the conserved mass monopole as well as the mass and current dipoles are in fact identical in both the harmonic and radiative constructions. In particular, we have $\dMbar=\dM$ for the constant (ADM) mass monopole, hence the slight abuse of notation in Eq.~\eqref{eq:xi1}.

The effect of this linear gauge transformation is to correct for the well-known logarithmic deviation of the retarded time in harmonic coordinates, with respect to the true space-time characteristic or light cone. After the change of gauge, the coordinate $u=t-r/c$ coincides (asymptotically when $r\to\infty$) with a null coordinate at the linearized level. The latter gauge transformation shifts the radiative metric away from harmonicity, since
\begin{equation}\label{eq:divrad1}
	\partial_\nu h^{\mu\nu}_{\text{rad}\, 1} = \Box \xi^\mu_{1} =
	\frac{2 \dM}{c^2 r^2}\,\eta^{0\mu}\,.
\end{equation}
Furthermore, one can easily show that, when $r\to \infty$ with $u=t-r/c$ held constant, the leading order $1/r$ in the metric is cancelled in the combination 
\begin{equation}\label{eq:kkhrad1}
	k_\mu k_\nu \,h^{\mu\nu}_{\text{rad}\, 1} = \mathcal{O}\left(\frac{1}{r^2}\right)\,,
\end{equation}
where $k^\mu = \eta^{\mu\nu}k_\nu = (1, \mathbf{n})$ denotes the outgoing Minkowskian null vector. 

Given any $n\geqslant 2$, let us recursively assume that: (i) we have obtained all the previous radiative PM coefficients $h^{\mu\nu}_{\text{rad}\, m}$ for any $m\leqslant n-1$; (ii) all of them admit an expansion as $r\to\infty$ with $u=t-r/c$ held constant in simple positive powers of $1/r$ (as opposed to a polylogarithmic behaviour); and (iii) all the previous coefficients satisfy the condition
\begin{equation}\label{eq:kkhradm}
\forall m\leqslant n-1\,,\qquad	k_\mu k_\nu \,h^{\mu\nu}_{\text{rad}\, m} =
	\mathcal{O}\left(\frac{1}{r^2}\right)\,.
\end{equation}
Note that the dominant piece when $r\to \infty$ (with $t-r/c=$ const) of the non-linear source term at the $n$-th order will be of order $1/r^2$ and will only be made of quadratic products of $h^{\mu\nu}_{\text{rad}\, m}$ (since each of the $h^{\mu\nu}_{\text{rad}\, m}$'s behaves like $1/r$). Under our recursive assumptions, in particular the induction hypothesis~\eqref{eq:kkhradm}, and from the structure of the source term at quadratic order, see \emph{e.g.} Eq.~(24) in~\cite{BlanchetLR}, one can prove that the $n$-th PM source term $\Lambda_{\text{rad}\, n}^{\mu\nu}\equiv\Lambda^{\mu\nu}[h_{\text{rad}\, m}; m\leqslant n-1]$ at leading order when $r\to\infty$ is of the form (see~\cite{B87} for details):
\begin{equation}\label{eq:Lambdan}
	\Lambda_{\text{rad}\, n}^{\mu\nu} = \frac{k^\mu
		k^\nu}{r^2}\,\sigma_{n}\bigl(t-r/c,\mathbf{n}\bigr) +
	\mathcal{O}\left(\frac{1}{r^3}\right)\,.
\end{equation}
This is the form of the stress-energy tensor of massless particles, \textit{i.e.}, gravitons in our case, with $\sigma_{n}$ being proportional to the $n$-th order contribution in the total power emitted by the massless waves. 

From Refs.~\cite{BD86, B87, BD92, B98quad} we know that logarithms in the asymptotic expansion when $r\to\infty$  only arise due to the retarded integral of source terms that behave like $1/r^2$. Hence the dominant term written in Eq.~\eqref{eq:Lambdan} is the only piece of $\Lambda_{\text{rad}\, n}^{\mu\nu}$ that can yield logarithms at order $n$. But now, thanks to the particular structure of this term, which follows from our recursive assumptions, we can gauge it away, thus constructing a coordinate system valid at the $n$-th PM order which avoids the appearance of logarithms. We find that an adequate gauge vector is
\begin{equation}\label{eq:xin}
	\xi^\mu_{n} = \FPprop\biggl[\left(\frac{r}{r_0}\right)^B
	\frac{c\, k^\mu}{2r^2} \int_{-\infty}^{t-r/c} \!\!\!\!\! \dd v \,
	\sigma_{n} (v,\mathbf{n})\biggr]\,.
\end{equation}
With this choice of gauge vector, the logarithms that will be generated from the gauge transformation will cancel the logarithms coming from the retarded integral of the source term~\eqref{eq:Lambdan}, see Ref.~\cite{B87} for more details. Hence, similarly to the corresponding steps~\eqref{eq:unharm}--\eqref{eq:vnharm} in the harmonic algorithm, we successively construct to the $n$-th order
\begin{equation}\label{eq:uwvnrad}
	\begin{aligned}
		u_{\text{rad}\, n}^{\mu\nu} 
		&=
		\FPprop\biggl[\left(\frac{r}{r_0}\right)^B\Lambda_{\text{rad}\, n}^{\mu\nu} \biggr]\,,\\
		w_{\text{rad}\, n}^{\mu} &= \partial_\nu u_{\text{rad}\, n}^{\mu\nu} = 
		\FPprop\biggl[B\left(\frac{r}{r_0}\right)^B\frac{n_i}{r}\,\Lambda_{\text{rad}\, n}^{\mu i} \biggr] \,,\\[0.3cm]
		v_{\text{rad}\, n}^{\mu\nu} &= \mathcal{V}^{\mu\nu}\bigl[w_{\text{rad}\, n}\bigr]\,,
	\end{aligned}
\end{equation}
so that the combination $u_{\text{rad}\, n}^{\mu\nu}+v_{\text{rad}\, n}^{\mu\nu}$ is divergenceless. Finally the $n$-th PM metric is defined by correcting for the new logarithms using the gauge transformation defined above:
\begin{align}\label{eq:hnrad}
	h_{\text{rad}\, n}^{\mu\nu} = u_{\text{rad}\, n}^{\mu\nu} + v_{\text{rad}\, n}^{\mu\nu} + \partial\xi_n^{\mu\nu}\,.
\end{align}
By construction, the radiative metric obeys the non-harmonic gauge condition
\begin{equation}\label{eq:dhnrad}
H_{\text{rad}\, n}^{\mu} \equiv	\partial_\mu h_{\text{rad}\, n}^{\mu\nu} = \Box \xi^\mu_{n} =
	\frac{c\,k^\mu}{2 r^2} \int_{-\infty}^{t-r/c} \!\!\dd v \, \sigma_{n}
	(v,\mathbf{n})\,,
\end{equation}
and the Einstein field equations, as given by \eqref{eq:EFE_vac}, are trivially satisfied to order $n$. The far-zone expansion of the full non-linear radiative metric constucted by virtue of this procedure is free of any logarithms, and the retarded time $u=t-r/c$ in these coordinates tends asymptotically toward a null coordinate at future null infinity. The metric, as a general functional of the moments $\{\dMbar_L,\overline{\dS}_L\}$, represents physically the most general solution to the vacuum field equations outside the isolated source.

\section{Relating the radiative and harmonic constructions}
\label{sec:General}

In Sec.~\ref{sec:HarmandRad}, we constructed two different metrics which both represent the most general solution of the vacuum field equations outside the matter source. We now want to relate them by imposing that they are physically equivalent, \emph{i.e.} differ only by a coordinate transformation and a multipole moment redefinition. We will then be able to explicitly express the canonical moments of the radiative algorithm $\{\dMbar_L,\overline{\dS}_L\}$ as functionals of the canonical moments of the harmonic algorithm $\{\dM_L,\dS_L\}$. This is the goal of this work, motivated by the fact that the computation of the tails-of-memory and spin-quadrupole-tails~\cite{TB23_ToM} has been performed using the radiative algorithm, while all previous results, \emph{i.e.} tails and tails-of-tails, and the rest of the computation of the 4PN flux, were achieved in harmonic coordinates. 

To determine the relation between $\{\dMbar_L,\overline{\dS}_L\}$ and $\{\dM_L,\dS_L\}$, we adapt the method of Ref.~\cite{MQ4PN_jauge} to the case where one of the two metrics does not satisfy the harmonic gauge condition. Recalling the transformation law of the gothic metric under a coordinate transformation $x^\mu \rightarrow x'^\mu$:
\begin{align}
\label{coordtrans}
h_\text{rad}^{\mu\nu}(x') =\frac{1}{|J|} \frac{\partial x'^\mu}{\partial x^\rho}  \frac{\partial x'^\nu}{\partial x^\sigma} \bigl( h_\text{harm}^{\rho\sigma}(x)+\eta^{\rho \sigma} \bigr) - \eta^{\mu\nu} \,,
\end{align}
where $J \equiv \det\left[ \partial x'^\mu/\partial x^\nu\right]$, we look for a coordinate shift $\varphi^\mu$ such that $x'^\mu = x^\mu + \varphi^\mu(x)$, then we have $\partial x'^\mu/\partial x^\rho = \delta^\mu_\rho + \partial_\rho \varphi^\mu(x)$ and $J = \det\left[\delta^\mu_\rho + \partial_\rho \varphi^\mu(x)\right]$.

By construction, both $h_\text{harm}^{\mu\nu}$ and $h_\text{rad}^{\mu\nu}$ admit a PM expansion as given respectively by~\eqref{eq:hharm} and~\eqref{eq:hrad}, which implies that the coordinate shift also admits the PM expansion
\begin{equation}
\varphi^\mu = \sum_{n=1}^{+\infty} G^n \varphi_n^\mu\,.
\end{equation}
Consistently, we assume that the respective canonical moments of the radiative and harmonic PM metrics also admit PM expansions,
\begin{equation}
\begin{aligned}
\dMbar_L &= \sum_{n=1}^{+\infty} G^{n-1}\dMbar_{n,L}\,, & \quad \dM_L &= \sum_{n=1}^{+\infty} G^{n-1} \dM_{n,L}\,,\\
\overline{\dS}_L &= \sum_{n=1}^{+\infty} G^{n-1}\overline{\dS}_{n,L}\,, & \quad \dS_L &= \sum_{n=1}^{+\infty} G^{n-1} \dS_{n,L}\,.
\end{aligned}
\end{equation}
As it is clear from the definition~\eqref{eq:hrad1}, the linear level is given by
\begin{equation}
h_{\text{rad}\,1}^{\mu\nu} = h_{\text{harm}\,1}^{\mu\nu} + \partial \varphi_1^{\mu\nu}\,,
 \end{equation}
where the coordinate shift is simply given by the gauge vector associated to the radiative construction, see~\eqref{eq:xi1}:
\begin{equation}
\varphi_1^\mu = \xi_1^\mu = \frac{2\dM}{c^2}\eta^{0\mu}\ln\left(\frac{r}{b_0}\right)\,,
\end{equation}
and where the moments are related by $\dMbar_L=\dM_L+ \mathcal{O}(G)$ and $\overline{\dS}_L=\dS_L+ \mathcal{O}(G)$. 

We then follow the reasoning of Ref.~\cite{MQ4PN_jauge}. We assume by induction that we have determined (for $n \geqslant 2$) the expressions of $\varphi^\mu$, $\dMbar_L$ and $\overline{\dS}_L$ as functionals of $\dM_L$ and $\dS_L$ up to $(n-1)$PM precision, where 1PM corresponds to the linear case treated above, \textit{i.e.}, that we have established the relations
\begin{subequations}
%\begin{equation}
\begin{align}
\varphi^\mu &= \sum_{m=1}^{n-1}G^m \Phi^\mu_m\left[\dM_K,\dS_K\right] + \mathcal{O}(G^{n})\,,\\
\dMbar_{L} &\equiv \sum_{m=1}^{n-1} G^{m-1} \mathcal{M}_{m,L}\left[\dM_K, \dS_K\right]+ \mathcal{O}(G^{n-1})\,,
\\
\overline{\dS}_{L} &\equiv \sum_{m=1}^{n-1} G^{m-1} \mathcal{S}_{m,L}\left[\dM_K, \dS_K\right]+ \mathcal{O}(G^{n-1})\,,
\end{align}
%\end{equation}
\end{subequations}
where $\Phi_m$, $\mathcal{M}_{m,L}$ and $\mathcal{S}_{m,L}$ are determined $m$-linear functionals of $\dM_L$ and $\dS_L$ for $m\leqslant n-1$. In principle, such functionals will be non-local in time at higher order, \textit{i.e.} involve some hereditary-like integrals. 

We then perform the expansion of Eq.~\eqref{coordtrans} up to order $n$PM, where we Taylor-expand the radiative metric to finite PM order, \textit{i.e.} using 
\begin{equation}
h_\text{rad}^{\mu\nu}(x') = h_\text{rad}^{\mu\nu}\bigl[x+\varphi(x)\bigr] =\sum_{m\geqslant 0}  \frac{1}{m!}\,\varphi^{\lambda_1} \cdots \varphi^{\lambda_m}\partial_{\lambda_1} \cdots \partial_{\lambda_m}h_\text{rad}^{\mu\nu}(x)\,.
\end{equation}
At the $n$PM order we find the relation
\begin{equation}
\label{eq:nPMcoordtrans}
h_{\text{rad}\,n}^{\mu\nu}(x)=h_{\text{harm}\,n}^{\mu\nu}(x)+ \partial \varphi_n^{\mu\nu}+\Omega_n^{\mu\nu}\bigl[h_{\text{harm}\, m}, \varphi_m; m\leqslant n-1\bigr]\,,
\end{equation} 
where $\Omega_n^{\mu\nu}$ is an explicitly known, non-linear and at least quadratic, functional of the coordinate shift and the harmonic metric at previous orders.\footnote{\label{footnote:explicit_expression_Omega}We recall from Ref.~\cite{MQ4PN_jauge} the expression of $\Omega^{\mu\nu}$ up to cubic order (in both $h^{\rho\sigma}$ and $\varphi^\rho$):
\begin{align*}
	\Omega^{\mu\nu}
	=
	& - \partial_\rho\left[\varphi^\rho\left(h^{\mu\nu}+\partial\varphi^{\mu\nu} + 2 h^{\sigma(\mu}\,\partial_\sigma\varphi^{\nu)} + \partial_\sigma\varphi^\mu\,\partial^\sigma\varphi^\nu \right)\right]\\
	& 
	+ 2\,\partial_\rho\varphi^{(\mu}\,h^{\nu)\rho}
	+ \partial^\rho\varphi^{\mu}\,\partial_\rho\varphi^{\nu} +  \frac{1}{2}\,\partial_{\rho\sigma}\left[\varphi^\rho\varphi^\sigma\left(h^{\mu\nu}+\partial\varphi^{\mu\nu}\right)\right] + h^{\rho\sigma}\,\partial_\rho\varphi^\mu\,\partial_\sigma\varphi^\nu\\
	& 
	+\frac{1}{2}\eta^{\mu\nu}\left[\partial_\rho\varphi^\sigma\partial_\sigma\varphi^\rho-(\partial_\rho\varphi^\rho)^2\right] - \frac{1}{3}\eta^{\mu\nu}\left[\partial_\rho\varphi^\sigma\partial_\sigma\varphi^\lambda\partial_\lambda\varphi^\rho - (\partial_\rho\varphi^\rho)^3\right]\\
	& - \frac{1}{2}\eta^{\mu\nu}\varphi^\rho\partial_\rho\left(\partial_\sigma\varphi^\lambda\partial_\lambda\varphi^\sigma-(\partial_\sigma\varphi^\sigma)^2\right) + \mathcal{O}(h^{4-p}\varphi^{p})\,.
\end{align*}
}

Thus, using the induction hypothesis, the only unknown term in Eq.~\eqref{eq:nPMcoordtrans} is the gauge shift $\varphi_n^{\mu}$, which we now determine. For this purpose, we define the quantity
\begin{equation}\label{eq:Deltadef}
\Delta_n^\mu \equiv - \partial_\nu \Omega_n^{\mu\nu}\,.
\end{equation}
Taking the divergence of~\eqref{eq:nPMcoordtrans}, and using the fact that $\partial_\nu\partial \theta^{\mu\nu} = \Box \theta^\mu$ for any~$\theta^\mu$ (recall our shorthand $\partial \theta^{\mu\nu}\equiv \partial^\mu \theta^\nu + \partial^\nu \theta^\mu - \eta^{\mu\nu}\partial_\rho \theta^{\rho}$), we find
\begin{equation}\label{eq:Deltares}
\Delta_n^\mu = \Box \bigl(\varphi_n^\mu - \xi_ n^\mu\bigr)\,.
\end{equation}

A difference with the treatment in Ref.~\cite{MQ4PN_jauge} is to be noted here. The analogue of the quantity $\Delta_n^\mu$ was defined in Ref.~\cite{MQ4PN_jauge} as $(\Delta_n^\mu)_\text{BFL} \equiv \Box \varphi_n^\mu$, but because of the harmonic gauge condition satisfied by the two metrics being related, this yielded  $(\Delta_n^\mu)_\text{BFL} = -\partial_\nu \Omega_n^{\mu\nu}$, \emph{i.e.} the same as our present definition~\eqref{eq:Deltadef}. When the harmonic gauge condition is relaxed it is important to proceed differently, starting from the \emph{definition}~\eqref{eq:Deltadef} and then deriving the result~\eqref{eq:Deltares}, where the gauge vector $\xi_ n^\mu$ given by~\eqref{eq:xin} is responsible for the non-harmonicity of the radiative metric, see~\eqref{eq:dhnrad}.

Next we apply the d'Alembertian operator on Eq.~\eqref{eq:nPMcoordtrans} to obtain
\begin{equation}\label{eq:sourceRelation}
\Lambda_{\text{rad}\,n}^{\mu\nu}= \Lambda_{\text{harm}\,n}^{\mu\nu}+\partial \Delta_n^{\mu\nu}+ \Box \Omega_n^{\mu\nu}\,,
\end{equation}
where we have used the fact that $\Box h_\text{rad}^{\mu\nu} = \Lambda_{\text{rad}\,n}^{\mu\nu}+\partial\Box \xi_n^{\mu\nu}$. To stick with the definition of the two algorithms we must now apply the inverse d'Alembertian operator with finite part prescription on Eq.~\eqref{eq:sourceRelation}, and arrive at
\begin{equation}\label{eq:uRelation0}
 u_{\text{rad}\,n}^{\mu\nu} = u_{\text{harm}\,n}^{\mu\nu} + \FPprop\biggl[\left(\frac{r}{r_0}\right)^B \partial \Delta_n^{\mu\nu}\biggl] + \FPprop\biggl[\left(\frac{r}{r_0}\right)^B \Box \Omega_n^{\mu\nu}\biggr] \,,
 \end{equation}
where the $u_{\text{harm}\,n}^{\mu\nu}$ is defined by Eq.~\eqref{eq:unharm} as laid out in the harmonic algorithm, while $u_{\text{rad}\,n}^{\mu\nu}$ is the equivalent quantity in the radiative algorithm given in Eqs.~\eqref{eq:uwvnrad}. Notice that in~\eqref{eq:uRelation0} we assumed that the Hadamard regularization scale $r_0$ is the same for the harmonic and radiative constructions.

Had the $\Box$ and $\partial$ operators commuted with the $\mathrm{FP}\Box^{-1}_\text{ret}$ operator, the previous equation would obviously simplify, but because of the presence of the regularization factor $(r/r_0)^B$, this is not the case, and we must introduce the \emph{a priori} non-zero ``commutators'' of these operators. So we can rewrite the previous equation as 
\begin{equation}\label{eq:uRelation}
 u_{\text{rad}\,n}^{\mu\nu}= u_{\text{harm}\,n}^{\mu\nu} + \partial \phi_n^{\mu\nu}+\Omega_n^{\mu\nu}+ X_n^{\mu\nu}+Y_n^{\mu\nu}\,,
\end{equation}
where we have introduced the two commutators
\begin{equation}
X_n^{\mu\nu} \equiv \Bigl[\mathrm{FP}\, \Box^{-1}_\mathrm{ret}, \,\Box\Bigr]
\,\Omega_n^{\mu\nu}\,,
\qquad Y_n^{\mu\nu} \equiv \Bigl[\mathrm{FP}\,
\Box^{-1}_\mathrm{ret}, \,\partial\Bigr]\,\Delta_n^{\mu\nu} \,,
\end{equation}
which can also be expressed in more details as~\cite{MQ4PN_jauge}
\begin{subequations}\label{eq:defXY}
\begin{align}	
X_n^{\mu\nu} &=  \FPprop \left[B \left(\frac{r}{r_0}\right)^B \left(- \frac{B+1}{r^2}\Omega_n^{\mu\nu}-\frac{2}{r}\partial_r \Omega_n^{\mu\nu} \right)\right]\,,\\
	Y_n^{\mu\nu} &= \FPprop \left[B \left(\frac{r}{r_0}\right)^B \frac{n^i}{r} \left(-\delta^{i\mu} \Delta_n^\nu - \delta^{i\nu}\Delta_n^\mu+\eta^{\mu\nu} \Delta_n^i \right)\right]\,.
\end{align}\end{subequations}
Note the very important presence of the explicit $B$ factor is the expressions of $X^{\mu\nu}_n$ and $Y^{\mu\nu}_n$, which will select the pole of the Laurent series when $B\to 0$. In addition to simplifying considerably the computations, this implies that we have $\Box X_n^{\mu\nu}=\Box Y_n^{\mu\nu}=0$. We also define
\begin{equation}\label{defphi}
\phi_n^\mu \equiv \FPprop\biggl[\left(\frac{r}{r_0}\right)^B \Delta_n^\mu\biggr]\,.
\end{equation}
Beware that this object is not the complete gauge transformation vector $\varphi_n^\mu$, as other contribution will show up later. 
 
To apply the harmonicity algorithm we compute the divergence of Eq.~\eqref{eq:uRelation}:
\begin{equation}\label{eq:wRelation}
w_{\text{rad}\,n}^\mu=w_{\text{harm}\,n}^\mu+\partial_\nu \bigl(X_n^{\mu\nu}+Y_n^{\mu\nu}\bigr)\,,
 \end{equation}
where we have used $\partial_\nu [ \partial \phi_n^{\mu\nu} ] = \Box \phi_n^\mu = \Delta_n^\mu$. We can check that $\Box w_{\text{rad}\,n}^{\mu}=0$, which is a necessary requirement to proceed with the MPM algorithm. Next we define $Z_n^{\mu\nu} \equiv \mathcal{V}^{\mu\nu}\left[W_n\right]$, where $\mathcal{V}^{\mu\nu}$ is the harmonicity algorithm applied to $W_n^\mu\equiv\partial_\nu (X_n^{\mu\nu}+Y_n^{\mu\nu})$, hence  $Z_n^{\mu\nu}$ is a solution to the vacuum equation whose divergence is exactly opposite to $W_n^\mu$, and we obtain
\begin{equation}\label{eq:vRelation}
v_{\text{rad}\,n}^{\mu\nu}=v_{\text{harm}\,n}^{\mu\nu}+Z_n^{\mu\nu}\,.
\end{equation}

Piecing it all together, we find that 
\begin{equation}\label{eq:calH}
h_{\text{rad}\,n}^{\mu\nu} = h_{\text{harm}\,n}^{\mu\nu}+\partial \phi_n^{\mu\nu}+\partial \xi_n^{\mu\nu}+\Omega_n^{\mu\nu}+ \mathcal{H}_n^{\mu\nu}\,,
\end{equation}
where $\mathcal{H}_n^{\mu\nu}\equiv X_n^{\mu\nu}+ Y_n^{\mu\nu}+ Z_n^{\mu\nu}$ is a divergenceless retarded homogeneous solution of the linearized Einstein vacuum equations, \emph{i.e.}, satisfying at once $\Box \mathcal{H}_n^{\mu\nu}\ = 0$ and $\partial_\nu \mathcal{H}_n^{\mu\nu} = 0$. In this respect, it can be uniquely parametrized~\cite{BD86} by two multipole ``source-type'' moments $\delta_n \dM_L$ and $\delta_n \dS_L$, which are functionals of $\{\dM_L, \dS_L\}$, along with a gauge vector $\zeta_n^\mu$ parametrized by four ``gauge-type'' moments, as in (37) of~\cite{BlanchetLR}.
 
Up to this step, we have assumed that we had related $\dMbar_{L}$ and $\dSbar_{L}$ to $\dM_L$ and $\dS_L$ to $(n-1)$PM precision. We can now redefine $\dMbar_{L}\rightarrow \dMbar_{L} - G^{n-1} \delta_n \dM_L$ and $\dSbar_{L}\rightarrow \dSbar_{L} - G^{n-1} \delta_n \dS_L$ so as to obtain this relationship up to $n$PM precision. This will not affect the result found in the recursion hypothesis since the correction is at $n$PM precision, but it will absorb the $\delta_n \dM_L$ and $\delta_n \dS_L$ moments into the linear approximation $h^{\mu\nu}_{\text{rad}\,1}$ of the radiative metric. Since we are correcting $\{\overline{\dM}_L, \overline{\dS}_L\}$ which parametrize the radiative metric in the left-hand-side of~\eqref{eq:calH}, and not $\{\dM_L, \dS_L\}$ in the right-hand-side, the expression for $\Omega^{\mu\nu}$, which only depends on $\{\dM_L, \dS_L\}$, does not need to be corrected.\footnote{A different yet totally equivalent method consists instead in redefining the linearized harmonic metric $h^{\mu\nu}_{\text{harm}\,1}$ in the right-hand side, namely by applying the redefinitions $\dM_{L}\rightarrow \dM_{L} + G^{n-1} \delta_n \dM_L$ and \mbox{$\dS_{L}\rightarrow \dS_{L} + G^{n-1} \delta_n \dS_L$}, but in that case we must include corrections due to the ``renormalization'' of $\Omega^{\mu\nu}$.} Finally, after this moment redefinition, we find that the two metrics are related by
\begin{equation}
h_{\text{rad}\,n}^{\mu\nu} = h_{\text{harm}\,n}^{\mu\nu}+\partial \varphi_n^{\mu\nu}+\Omega_n^{\mu\nu}\,,
\end{equation}
in which we have finally explicitly determined the looked-for gauge vector as
\begin{equation}
\label{eq:construction_varphi}
	\varphi_n^\mu \equiv \phi_n^\mu + \xi_n^\mu + \zeta_n^\mu\,.
\end{equation}
By construction, this vector is a functional of $\dM_L$ and $\dS_L$, so we can write
\begin{subequations}
\begin{equation}
\varphi^\mu = \sum_{m=1}^{n}G^m \Phi_m^\mu\left[\dM_K,\dS_K\right] + \mathcal{O}(G^{n+1})\,.
\end{equation}
Simarly, the corrections $\delta_n \dM_L \equiv G^{n-1} \mathcal{M}_{n,L}[\dM_K,\dS_K]$ and $\delta_n \dS_L \equiv G^{n-1} \mathcal{S}_{n,L}[\dM_K,\dS_K]$ are also functionals of $\dM_L$ and $\dS_L$, and we find
\begin{align}
\dMbar_{L} &\equiv \sum_{m=1}^{n} G^{m-1} \mathcal{M}_{m,L}\left[\dM_K, \dS_K\right]+ \mathcal{O}(G^{n})\,,
\\
\overline{\dS}_{L} &\equiv \sum_{m=1}^{n} G^{m-1} \mathcal{S}_{m,L}\left[\dM_K, \dS_K\right]+ \mathcal{O}(G^{n})\,.
\end{align}
\end{subequations}
This completes the recursion procedure, and we have thus proven that we can explicitly construct $\varphi^\mu$, $\dMbar_L$ and $\overline{\dS}_L$ as functionals of $\dM_L$ and $\dS_L$ to any finite PM order.

With those results in hand, and the help of the integration formulae presented in App.~\ref{sec:Retard}, we can implement all the terms up to cubic non-linear order, using the \emph{xAct} library of the \emph{Mathematica} software~\cite{xtensor}. In particular, this computation lead to the result at~4PN~order given in Eq.~\eqref{eq:deltaMijres}, and was vital in the calculation of \emph{tail-of-memory} and \mbox{\emph{spin-quadrupole tail}} terms~\cite{TB23_ToM}.

\section{Application to non-linear tail interactions}
\label{sec:Tails}

\subsection{Quadratic tails}

We employ the method exposed in Sec.~\ref{sec:General} to explicitly construct the relation between the radiative and harmonic metrics at quadratic order, to all relevant orders in $1/r$. To do so, we will focus on the tail effect, which arises due to the interaction between the static ADM mass $\dM$ and the various dynamical mass and current multipole moments, thus generating features in the waveform that are non-local in time. As it is the goal of our work, we focus on the corrections to the mass quadrupole moment $\dM_{ij}$, but the computations presented hereafter are easily generalized to the tails of other moments.
 
Recall that at 4PN order, other quadratic interactions enter the relation between the quadrupole moment $\dM_{ij}$ and the observable at infinity (dubbed ``radiative moment''), such as the memory type interaction $\dM_{ij}\times \dM_{ij}$ that enters the metric at 2.5PN order. Nevertheless, here we are looking for the relation between the two quadrupole moments $\dM_{ij}$ and $\dMbar_{ij}$ in the two algorithms, and this relation at quadratic order must involve the mass $\dM$ which parametrizes the linearized gauge vector~\eqref{eq:xi1}. It follows that the only quadratic multipole interaction which can contribute to the relation between the two moments is $\dM\times \dM_{ij}$, and we shall find that this relation is purely instantaneous in this case.

In order to obtain the relation between the harmonic and radiative metrics, and thus the associated correction to the quadrupole moments, we will in fact use two independent methods: (i) the one exposed in Sec.~\ref{sec:General}, which does not require knowing the full expressions of the quadratic metrics; and (ii) the explicit computation and comparison of the full harmonic and radiative metrics. We will find exactly the same result, thus confirming the soundness of the method exposed in Sec.~\ref{sec:General}.

Let us first derive the complete metrics for the tail interaction in the two harmonic and radiative algorithms, and read off the quadrupole moments from these two metric in both coordinates. We shall check that the difference between those two quadrupole moments perfectly match the prediction following from the general procedure in Sec.~\ref{sec:General}. The tail sector of the metric in the harmonic algorithm (defined in Sec.~\ref{sec:harmalg}) is given in~App.~B~of~\cite{BD92}, and reads\footnote{For the sake of lightness, we set $c=1$ throughout Sec.~\ref{sec:Tails}.}
\begin{subequations}\label{eq:htailharm}
\begin{align}
h_{\text{harm}\,\dM\times \dM_{ij}}^{00} &= 8 \dM  n_{ab}   \int_1^{+\infty} \dd x \,Q_2(x) \dM^{(4)}_{ab}(t-rx) \nonumber\\
&\quad+ \frac{\dM n_{ab}}{r}\Big(10  \dM^{(3)}_{ab} + 7 r^{-1}  \dM^{(2)}_{ab}- 21 r^{-2}  \dM^{(1)}_{ab} -21 r^{-3}\dM^{}_{ab}\Big)\,,\\
%%%%%%%%%%%%%%%%%%%%%%%%%%%%%%%%%%%%%%%%%%%%%%%%%%%%%%%%%%%%%%%%%%%%%%%%%%%%%%%%
h_{\text{harm}\,\dM\times \dM_{ij}}^{0i} &= 8 \dM  n_{a}   \int_1^{+\infty} \dd x \,Q_1(x) \dM^{(4)}_{ai}(t-rx)\nonumber\\
&\quad +\frac{\dM n_{iab}}{r}\bigg[-\frac{1}{3}\dM^{(3)}_{ab}- r^{-1}\dM^{(2)}_{ab}- r^{-2}\dM^{(1)}_{ab} \bigg]\nonumber\\ 
&\quad+ \frac{\dM n_{a}}{r}\bigg[\frac{19}{3}\dM^{(3)}_{ai}-5 r^{-1}\dM^{(2)}_{ai}- 5r^{-2}\dM^{(1)}_{ai} \bigg]\,,\\
%%%%%%%%%%%%%%%%%%%%%%%%%%%%%%%%%%%%%%%%%%%%%%%%%%%%%%%%%%%%%%%%%%%%%%%%%%%%%%%%
h_{\text{harm}\,\dM\times \dM_{ij}}^{ij} &= 8 \dM \int_1^{+\infty} \dd x \,Q_0(x) \dM^{(4)}_{ij}(t-rx)\nonumber\\
&\quad +\frac{\dM n_{ijab}}{r}\bigg[ -\frac{1}{2}\dM^{(3)}_{ab}-3r^{-1}\dM^{(2)}_{ab}-\frac{15}{2} r^{-2}\dM^{(1)}_{ab}-\frac{15}{2} r^{-3}\dM_{ab}\bigg]\nonumber\\
&\quad +\frac{\dM \delta_{ij}n_{ab}}{r}\bigg[ -\frac{11}{6}\dM^{(3)}_{ab}-2r^{-1}\dM^{(2)}_{ab}-\frac{1}{2} r^{-2}\dM^{(1)}_{ab}-\frac{1}{2} r^{-3}\dM_{ab}\bigg]
\nonumber\\
&\quad +\frac{\dM n_{a(i}}{r}\bigg[4\dM^{(3)}_{j)a}+6 r^{-1}\dM^{(2)}_{j)a}+6 r^{-2}\dM^{(1)}_{j)a}+6 r^{-3}\dM_{j)a}\bigg]\nonumber\\
&\quad +\frac{\dM }{r}\bigg[ -\frac{11}{3}\dM^{(3)}_{ij}-4r^{-1}\dM^{(2)}_{ij}-r^{-2}\dM^{(1)}_{ij}-r^{-3}\dM_{ij}\bigg]\,,
\end{align}
\end{subequations}
where $Q_m(x)$ denotes the Legendre function of the second kind, here defined with a branch cut from $-\infty$ to 1, and given explicitly in terms of the Legendre polynomial $P_m(x)$ by
\begin{equation}\label{eq:Qmdef}
Q_{m} (x) =  \frac{1}{2} P_m (x) \, \ln
		\left(\frac{x+1}{x-1} \right)- \sum^{m}_{ j=1} \frac{1}{j}
		P_{m-j}(x) P_{j-1}(x)\,.
\end{equation}

On the other hand, we have applied the radiative construction of Ref.~\cite{B87} for this same particular tail interaction, thus following the radiative algorithm described in Sec.~\ref{sec:radalg}. The moments involved in the construction of the tail sector of the radiative metric are $\dM$ and $\dMbar_{ij}$ in this construction (recall that $\dM=\dMbar$). Following our convention to denote the coordinates by dummy variables $(t, r, \mathbf{n})$, which correspond here to the radiative coordinate system which defines the radiative metric, we find that the tail metric in the radiative algorithm is explicitly given by 
\begin{subequations}\label{eq:rad_metric_full}
\begin{align}
h_{\text{rad}\,\dM\times \dMbar_{ij}}^{00} &= 8 \dM  n_{ab}   \int_1^{+\infty} \dd x \,\overline{Q}_2(x,r)\,\dMbar^{(4)}_{ab}(t-rx)\nonumber\\
&\quad+ \frac{\dM n_{ab}}{r}\Big(\frac{118}{15}  \dMbar^{(3)}_{ab} + \frac{23}{5} r^{-1} \dMbar^{(2)}_{ab}- \frac{117}{5} r^{-2}  \dMbar^{(1)}_{ab} -21 r^{-3}  \dMbar_{ab}\Big)\,,\\
%%%%%%%%%%%%%%%%%%%%%%%%%%%%%%%%%%%%%%%%%%%%%%%%%%%%%%%%%%%%%%%%%%%%%%%%%%%%%%%%
h_{\text{rad}\,\dM\times \dMbar_{ij}}^{0i} &= 8 \dM  n_{a}   \int_1^{+\infty} \dd x \,\overline{Q}_1(x,r)\, \dMbar^{(4)}_{ai}(t-rx)\nonumber\\
&\quad +\frac{\dM n_{iab}}{r}\bigg[\dMbar^{(3)}_{ab}+3 r^{-1}\dMbar^{(2)}_{ab}- r^{-2}\dMbar^{(1)}_{ab} \bigg]\nonumber\\
&\quad+ \frac{\dM n_{a}}{r}\bigg[\frac{43}{15}\dMbar^{(3)}_{ai}-\frac{107}{15} r^{-1}\dMbar^{(2)}_{ai}- 5r^{-2}\dMbar^{(1)}_{ai} \bigg]\,,\\
%%%%%%%%%%%%%%%%%%%%%%%%%%%%%%%%%%%%%%%%%%%%%%%%%%%%%%%%%%%%%%%%%%%%%%%%%%%%%%%%
h_{\text{rad}\,\dM\times \dMbar_{ij}}^{ij} &= 8 \dM \int_1^{+\infty} \dd x \,\overline{Q}_0(x,r)\, \dMbar^{(4)}_{ij}(t-rx) \nonumber\\
&\quad +\frac{\dM n_{ijab}}{r}\bigg[ -\frac{1}{2}\dMbar^{(3)}_{ab}-3r^{-1}\dMbar^{(2)}_{ab}-\frac{15}{2} r^{-2}\dMbar^{(1)}_{ab}-\frac{15}{2} r^{-3}\dMbar_{ab}\bigg]\nonumber\\
&\quad +\frac{\dM \delta_{ij} n_{ab}}{r}\bigg[ -\frac{1}{2}\dMbar^{(3)}_{ab}-2r^{-1}\dMbar^{(2)}_{ab}-\frac{1}{2} r^{-2}\dMbar^{(1)}_{ab}-\frac{1}{2} r^{-3}\dMbar_{ab}\bigg]
\nonumber\\
&\quad +\frac{\dM n_{a(i}}{r}\bigg[4\dMbar^{(3)}_{j)a}+6 r^{-1}\dMbar^{(2)}_{j)a}+6 r^{-2}\dMbar^{(1)}_{j)a}+6 r^{-3}\dMbar_{j)a}\bigg]\nonumber\\
&\quad +\frac{\dM }{r}\bigg[ -\frac{107}{15}\dMbar^{(3)}_{ij}-4r^{-1}\dMbar^{(2)}_{ij}-r^{-2}\dMbar^{(1)}_{ij}-r^{-3}\dMbar_{ij}\bigg]\,.
\end{align}
\end{subequations}
The main difference between the two expressions~\eqref{eq:htailharm} and~\eqref{eq:rad_metric_full} of the metric is that the tail terms are given by integrals over the legendre function $Q_m(x)$ in the harmonic case while the tails in the radiative case are given by integrals over the particular combination
\begin{equation}
	\label{eq:hatQm}
	\overline{Q}_m(x,r) \equiv Q_m(x)-\frac{1}{2}P_m(x)\ln\left(\frac{r}{r_0}\right)\,.
\end{equation}

It is easy to see why the tail terms in the harmonic metric involving $Q_m(x)$ generate logarithms in the far-zone expansion $r\to\infty$, while the particular combination $\overline{Q}_m(x,r)$ defined by~\eqref{eq:hatQm} does not produce any logarithms of $r$ in the radiative metric, although the tail terms are still given by a ``hereditary'' integral, of course. Posing $\tau=r(x-1)$ and $u=t-r$ the tail integral reads
\begin{equation}
 \int_1^{+\infty} \dd x \,Q_m(x) F(t-r x) = \frac{1}{r} \int_0^{+\infty} \dd \tau \,Q_m\Bigl(1+\frac{\tau}{r}\Bigr) F(u-\tau) \,,
\end{equation}
where the function $F(t)$ is a time derivative of the multipole moment and we recall that it vanishes identically for $t< -\mathcal{T}$ (stationarity in the past). Hence, as shown in~\cite{BD92, B98tail} the far-zone limit $r \to+\infty$ (with $u\equiv t-r$ held constant) of the tail integral can be obtained by inserting the expansion of the Legendre function when $x\to 1^+$. Using the expression~\eqref{eq:Qmdef} for the Legendre functions and expanding around  $x\to 1^+$, we obtain the following formal expansion series when $r \to +\infty$,
\begin{subequations}
\begin{align}
	Q_m& \Bigl(1+\frac{\tau}{r}\Bigr) =  - \frac{1}{2} P_m \Bigl(1+\frac{\tau}{r}\Bigr) \, \ln
	\left(\frac{\tau}{2r} \right) +  \sum_{ k=0}^{+\infty} \lambda^{m}_{k} \,\left(\frac{\tau}{r}\right)^k\,,\label{eq:Qmlogr}\\
\text{where}\quad\lambda^{m}_{k} &= - \!\!\sum_{i=0}^{\min\left(m,\,k-1\right)} \frac{(-)^{k-i}(m+i)!}{2^{k+1}(k-i)(m-i)!(i!)^2} \nonumber\\& ~\quad - \sum_{j=k+1}^{m} \sum_{i=0}^k\frac{(m+2i-j)!(k+j-2i-1)!}{2^k (j-i) (m-j)!(j-k-1)!(i!)^2[(k-i)!]^2} \,,
\end{align}
\end{subequations}
and where the dependencies in $\ln r$ are contained in the first term of~\eqref{eq:Qmlogr}. We can see that the first term in the expression of $\lambda^{m}_{k}$ vanishes for $k=0$ (yielding \mbox{$\lambda^{m}_{0} = - H_m$} where $H_m$ is the usual harmonic number) and the second term vanishes for \mbox{$k\geqslant m$}. 

Conversely, the tail terms in radiative coordinates written in terms of the combination $\overline{Q}_m(x,r)$ do not produce any logarithms in the far-zone expansion. Indeed we have
\begin{equation}\label{eq:inttailradiatif}
	\int_1^{+\infty} \dd x\, \overline{Q}_m(x,r)\,F(t-rx) = \frac{1}{r} \int_0^{+\infty} \dd \tau\, \Bigl[R_m\Bigl(1+\frac{\tau}{r}\Bigr)-\frac{1}{2}P_m\Bigl(1+\frac{\tau}{r}\Bigr) \ln\Bigl(\frac{\tau}{2 r_0}\Bigr)\Bigr]F(u-\tau)  \,,
\end{equation}
where $u\equiv t-r$ is the retarded time of the radiative coordinates, and where we have defined
\begin{equation}
R_m(x) \equiv Q_m(x) +~\frac{1}{2}P_m(x)\ln\left(\frac{x-1}{2}\right) \,,
\end{equation}
which is regular when $x\rightarrow 1^+$.  Since the integrand of~\eqref{eq:inttailradiatif} is regular as $r\rightarrow \infty$, this integral's expansion for large $r$ is free of any logarithmic terms. Furthermore, we find that the first term in~\eqref{eq:inttailradiatif} admits the following asymptotic expansion in simple powers of $1/r$, 
\begin{equation}\label{eq:Rmreg}
	\int_0^{+\infty} \dd \tau\, R_m\Bigl(1+\frac{\tau}{r}\Bigr) F(u-\tau) = \sum_{ k=0}^{+\infty} k! \,\lambda^{m}_{k} \,\frac{F^{(-k-1)}(u)}{r^{k}}\,.
\end{equation}

All of this is illustrated by the expression of the leading far-zone limit of Eqs.~\eqref{eq:htailharm} and~\eqref{eq:rad_metric_full}. In harmonic coordinates we have
\begin{subequations}
	\begin{align}
		h_{\text{harm}\,\dM\times \dM_{ij}}^{00} = & - \frac{4\dM n_{ab}}{r} \int_0^{+\infty} \dd \tau\, \bigg[\ln\left(\frac{\tau}{2r}\right) + \frac{1}{2}\bigg]\dM^{(4)}_{ab}(u-\tau)+ \mathcal{O}\left(\frac{\ln r}{r^{2}}\right)\,,\\
		h_{\text{harm}\,\dM\times \dM_{ij}}^{0i} = & - \frac{4\dM n_{a}}{r} \int_0^{+\infty} \dd \tau\, \bigg[\ln\left(\frac{\tau}{2r}\right) + \frac{5}{12}\bigg]\dM^{(4)}_{ia}(u-\tau) \nonumber\\
		&- \frac{\dM}{3r} n_{iab}\dM^{(3)}_{ab}+ \mathcal{O}\left(\frac{\ln r}{r^{2}}\right)\,,\\
		h_{\text{harm}\,\dM\times \dM_{ij}}^{ij} = & - \frac{4\dM }{r} \int_0^{+\infty} \dd \tau\, \bigg[\ln\left(\frac{\tau}{2r}\right) + \frac{11}{12}\bigg]\dM^{(4)}_{ij}(u-\tau)  \nonumber\\
		&+ \frac{\dM}{r}\left[- \frac{1}{2}n_{ijab} \dM^{(3)}_{ab}+ 4 n_{a(i} \dM^{(3)}_{j)a}- \frac{11}{6}\delta_{ij}n_{ab} \dM^{(3)}_{ab}\right] + \mathcal{O}\left(\frac{\ln r}{r^{2}}\right)\,,
	\end{align}
\end{subequations}
which clearly exhibits the usual far-zone logarithms associated with harmonic coordinates. By contrast the leading far-zone behaviour of the radiative metric reads
\begin{subequations}
\begin{align}
h_{\text{rad}\,\dM\times \dMbar_{ij}}^{00} = &  - \frac{4\dM n_{ab}}{r} \int_0^{+\infty} \dd \tau\, \bigg[\ln\left(\frac{\tau}{2r_0}\right) + \frac{31}{30}\bigg]\,\dMbar^{(4)}_{ab}(u-\tau)+ \mathcal{O}\left(\frac{1}{r^{2}}\right)\,,\\
%%%%%%%%%%%%%%%%%%%%%%%%%%%%%%%%%%%%%%%%%%%%%%%%%%%%%%%%%%%%%
h_{\text{rad}\,\dM\times \dMbar_{ij}}^{0i} = & - \frac{4\dM n_{a}}{r} \int_0^{+\infty} \dd \tau\, \bigg[\ln\left(\frac{\tau}{2r_0}\right) + \frac{77}{60}\bigg]\dMbar^{(4)}_{ia}(u-\tau) \nonumber\\
&+ \frac{\dM}{r}n_{iab}\dMbar^{(3)}_{ab}+ \mathcal{O}\left(\frac{1}{r^{2}}\right)\,,\\
%%%%%%%%%%%%%%%%%%%%%%%%%%%%%%%%%%%%%%%%%%%%%%%%%%%%%%%%%%%%%
h_{\text{rad}\,\dM\times \dMbar_{ij}}^{ij} = & - \frac{4\dM}{r} \int_0^{+\infty} \dd \tau\, \bigg[\ln\left(\frac{\tau}{2r_0}\right) + \frac{107}{60}\bigg]\dMbar^{(4)}_{ij}(u-\tau) \nonumber\\
&+ \frac{\dM}{r}\left[ - \frac{1}{2} n_{ijab} \dMbar^{(3)}_{ab}+ 4 n_{a(i} \dMbar^{(3)}_{j)a}- \frac{1}{2}\delta_{ij} n_{ab} \dMbar^{(3)}_{ab}\right]+ \mathcal{O}\left(\frac{1}{r^{2}}\right) \,,
\end{align}
\end{subequations}
hence the expansion of the tail term is now free of logarithms, although it now depends on the Hadamard regularization scale $r_0$.

Now that we have both explicit metrics including tail terms~\eqref{eq:htailharm} and~\eqref{eq:rad_metric_full} at our disposal, we can check that they are indeed physically equivalent to this order. This means that the two metrics should differ by a non-linear coordinate transformation together with the redefinition of the quadrupole moment $\dM_{ij}\longrightarrow\dMbar_{ij}$, \emph{i.e.} that
\begin{equation}
	\label{eq:equivalenceEquation}
	h_{\text{rad}\, \dMbar_{ij}}^{\mu\nu}+G \,h_{\text{rad}\,\dM\times \dMbar_{ij}}^{\mu\nu}
	= h_{\text{harm}\, \dM_{ij}}^{\mu\nu} + G \Bigl[ h_{\text{harm}\,\dM\times \dM_{ij}}^{\mu\nu} + \partial \varphi_{\dM\times \dM_{ij}}^{\mu\nu} + \Omega_{\dM\times\dM_{ij}}^{\mu\nu} \Bigr] + \mathcal{O}(G^2)\,,
\end{equation}
where the radiative metric in the left-hand side is defined with $\dMbar_{ij}$ and the right-hand side is defined with $\dM_{ij}$, and with a linear gauge transformation vector $\varphi_{\dM\times \dM_{ij}}^\mu$ and $\Omega_{\dM\times \dM_{ij}}^{\mu\nu}$ the non-linear part of the coordinate transformation. Moreover, recall that at the linear level, there is only a mass monopole contribution to the gauge vector, $\varphi_{\dM}^\mu\equiv\xi_{\dM}^\mu$ as given by~\eqref{eq:xi1}, and in particular $\varphi_{\dM_{ij}}^\mu$ is vanishing.

Applying the general procedure of Sec.~\ref{sec:General}, we find that the relevant gauge vector corresponding to the tail interaction, given explicitly by \eqref{eq:construction_varphi}, reads
\begin{subequations}\label{eq:varphiexpl}
	\begin{align}
	\varphi_{\dM\times \dM_{ij}}^0 = - \frac{4\dM}{3r} n_{ab}\dM^{(2)}_{ab}\,,\qquad \varphi_{\dM\times \dM_{ij}}^i = 0	\,,  
	\end{align}
whereas the non-linear correction term, defined by Footnote \ref{footnote:explicit_expression_Omega}, reads
\begin{align}
	\Omega_{\dM\times \dM_{ij}}^{00} &= -4\dM  \partial_{ab}\left[r^{-1}\ln\left(\frac{r}{b_0}\right)\dM^{(1)}_{ab}\right]- \frac{8\dM}{r^3}n_{ab}\dM^{(1)}_{ab}\,, \\
	\Omega_{\dM\times \dM_{ij}}^{0i} &= 4 \dM  \partial_a \left[r^{-1} \ln\left(\frac{r}{b_0}\right) \dM^{(2)}_{ai}\right]\,, \\
	\Omega_{\dM\times \dM_{ij}}^{ij} &= - \frac{4\dM}{r}\ln\left(\frac{r}{b_0}\right)\dM^{(3)}_{ij}\,.
\end{align}
\end{subequations}
Most importantly, after computing the explicit expression of $\mathcal{H}_n^{\mu\nu}$ using the general method described in Sec.~\ref{sec:General}, we determine that this coordinate transformation must be associated with the following redefinition of the mass quadrupole moment:
\begin{equation}
	\label{eq:dMij_tail}
	\dMbar_{ij} =
	\dM_{ij}+ G\,\dM\left[-\frac{26}{15}+2 \ln\left(\frac{r_0}{b_0}\right)\right] \dM_{ij}^{(1)} + \mathcal{O}(G^2)\,.
\end{equation}
It is straightforward to show that the relation~\eqref{eq:equivalenceEquation} is satisfied, to all orders in $1/r$ and for $\varphi^\mu_{\dM\times \dM_{ij}}$ and $\Omega^{\mu\nu}_{\dM\times \dM_{ij}}$ defined by Eqs. \eqref{eq:varphiexpl}, if and only if the moment redefinition given by~\eqref{eq:dMij_tail} holds, hence confirming the soundness of the method of Sec.~\ref{sec:General}.

We can also discuss this equivalence, in a simpler way, directly at the level of the radiative quadrupole moment $\mathcal{U}_{ij}$ defined at future null infinity $\mathcal{I}^{+}$ in both constructions. Note that in previous works on tails and iterated tails in \emph{harmonic} coordinates~\cite{BD92, B98tail, FBI15, MBF16}, it was shown that the leading logarithms present in the asymptotic waveform (following the algorithm of Sec.~\ref{sec:harmalg}), can be removed by just the linear coordinate transformation $x'^\mu = x^\mu + G \,\xi_1^\mu$, where~$\xi_1^\mu$ is given by~\eqref{eq:xi1}. In the case of quadratic tails, this \textit{ad hoc} procedure, followed by a transverse-traceless (TT) projection of the spatial metric, yields the radiative moment:
\begin{align}\label{eq:Uijharm}
	\mathcal{U}_{ij}(u) = \dM^{(2)}_{ij}(u) + 2G\dM \int_0^{+\infty} \dd \tau\, \bigg[ \ln\left(\frac{\tau}{2b_0}\right) + \frac{11}{12}\bigg] \dM^{(4)}_{ij}(u-\tau)+ \mathcal{O}(G^2)\,.
\end{align}
Here $\dM_{ij}$ is the canonical moment associated to the harmonic-coordinate construction, and again we use the dummy notation $u=t-r$ for the retarded time in radiative coordinates. Notice that the effect of this coordinate transformation is to replace the logarithm $\ln r$ in harmonic coordinates by the constant $\ln b_0$.

On the other hand, when following the procedure for the radiative construction (namely the algorithm of Sec.~\ref{sec:radalg}), we find that the radiative moment reads
\begin{align}\label{eq:Uijrad}
	\mathcal{U}_{ij}(u) = \dMbar^{(2)}_{ij}(u) + 2G\dM \int_0^{+\infty} \dd \tau\, \bigg[ \ln\left(\frac{\tau}{2r_0}\right) + \frac{107}{60}\bigg] \dMbar^{(4)}_{ij}(u-\tau)+ \mathcal{O}(G^2)\,.
\end{align}
Of course this object is the same as in~\eqref{eq:Uijharm}, however it is expressed in terms of the canonical moment $\dMbar_{ij}$ associated to the radiative algorithm. As one can immediately check, the relation~\eqref{eq:dMij_tail} we have found between the two canonical moments exactly reconciles the results~\eqref{eq:Uijharm} and~\eqref{eq:Uijrad}, which again confirms our method in Sec.~\ref{sec:General}.

\subsection{Cubic tails-of-tails}

We have pushed these calculations to cubic order to include the tail-of-tail effect, due to the non-linear interaction $\dM \times \dM \times \dM_{ij}$, extending~\eqref{eq:Uijharm} and~\eqref{eq:Uijrad} to the next order in~$G$. In the harmonic algorithm, after the suitable coordinate transformation to get rid of the logarithms, we find~\cite{B98tail, FBI15}
\begin{align}\label{eq:Uijharm2}
	\mathcal{U}_{ij} =&\, \dM^{(2)}_{ij} + 2G\dM \int_0^{+\infty} \dd \tau\, \bigg[ \ln\left(\frac{\tau}{2b_0}\right) + \frac{11}{12}\bigg] \dM^{(4)}_{ij}(u-\tau) \nonumber\\ &+ 2G^2\dM^2 \int_0^{+\infty} \dd \tau\, \bigg[ \ln^2\left(\frac{\tau}{2b_0}\right) + \frac{11}{6} \ln\left(\frac{\tau}{2b_0}\right) - \frac{107}{105} \ln\left(\frac{\tau}{2r_0}\right) + \frac{124627}{44100}\bigg] \dM^{(5)}_{ij}(u-\tau) \nonumber\\&+ \mathcal{O}(G^3)\,,
\end{align}
Note that this result involves two arbitrary scales, which are important to distinguish: the scale $b_0$ which enters the asymptotic coordinate transformation $u \longrightarrow u - 2G\dM \,\ln(r/b_0)$, and, at cubic order, the Hadamard regularization scale $r_0$. It is known~\cite{GRoss10, GRR12} that $r_0$ can be interpreted as a renormalization scale and its running obeys a renormalization group equation. Indeed, we recognize  the coefficient in front of the $\ln(r_0)$ in \eqref{eq:Uijharm2}: it is exactly the beta-function coefficient associated to the renormalization of the mass quadrupole moment, given to be $\beta_2 =-\frac{214}{105}$ in (45) of \cite{GRoss10}.

%The coefficient $\beta_{\dM_{ij}}=-\frac{214}{105}$ in front of the $\ln r_0$ is the beta-function coefficient corresponding to the renormalization of the mass quadrupole moment. 

We have redone the calculation of tails-of-tails using the radiative algorithm described in Sec.~\ref{sec:radalg}. In this case, as already noticed, the result depends only on the regularization scale $r_0$ but is to be expressed in terms of the quadrupole moment $\dMbar_{ij}$ associated to the radiative algorithm. We obtain
\begin{align}\label{eq:Uijrad2}
	\mathcal{U}_{ij} &= \dMbar^{(2)}_{ij} + 2G\dM \int_0^{+\infty} \dd \tau\, \bigg[ \ln\left(\frac{\tau}{2r_0}\right) + \frac{107}{60}\bigg] \dMbar^{(4)}_{ij}(u-\tau) \nonumber\\ &+ 2G^2\dM^2 \int_0^{+\infty} \dd \tau\, \bigg[ \ln^2\left(\frac{\tau}{2r_0}\right) + \frac{107}{42} \ln\left(\frac{\tau}{2r_0}\right) + \frac{40\,037 }{8820}\bigg] \dMbar^{(5)}_{ij}(u-\tau) + \mathcal{O}(G^3)\,.
\end{align}
Finally by employing the method of Sec.~\ref{sec:General} up to cubic order for tails and tails-of-tails we have obtained the relationship between the quadrupoles in the two constructions as
\begin{equation}\label{eq:deltaMijtailsoftails}
	\begin{aligned}
		\dMbar_{ij} =
		\dM_{ij}
		&+ G\,\dM \left[-\frac{26}{15}+2 \ln\left(\frac{r_0}{b_0}\right)\right] \dM_{ij}^{(1)}\\
		&+
		G^2 \,\dM^2 \!\left[\frac{124}{45}-\frac{52}{15} \ln\left(\frac{r_0}{b_0}\right) +2  \ln^2\left(\frac{r_0}{b_0}\right)  \right] \dM_{ij}^{(2)} + \mathcal{O}(G^3)\,.
	\end{aligned}
\end{equation}

This result is indeed the unique relationship between the moments that reconciles the two results~\eqref{eq:Uijharm2} and~\eqref{eq:Uijrad2}. Notice that the constant scale~$b_0$ in~\eqref{eq:deltaMijtailsoftails} was introduced ``automatically'' in the linear gauge transformation between the harmonic and radiative linear metrics, see~\eqref{eq:hrad1}--\eqref{eq:xi1}. This scale is identical to the one introduced ``by hand'' in harmonic coordinates, see Eq.~(3.1) in~\cite{FBI15} or (4.2) in~\cite{MBF16}.

As discussed in Sec.~\ref{sec:motivations}, the previous \emph{ad hoc} method for removing the logarithms in the harmonic metric was satisfactory for tails and iterated tail interactions since the coordinate transformation could easily be guessed. However, when considering more complicated non-linear interactions such as the tails-of-memory occuring at 4PN order, the coordinate transformation is more difficult to implement, and it is more convenient to switch to the radiative algorithm since it directly constructs the metric in radiative coordinates and automatically removes the logarithms. The price we have to pay is that,  in the end, we must apply the correction $\{\dMbar_L,\overline{\dS}_L\}\longrightarrow\{\dM_L,\dS_L\}$ in order to match with previous results derived in harmonic coordinates, and most importantly the explicit expressions of $\{\dM_L,\dS_L\}$ as functions of the source at 4PN order~\cite{MQ4PN_jauge}.

\section{Cubic tail-of-memory interactions at 4PN order}
\label{sec:cubicRes}

We now apply our method to the cubic ``tails-of-memory'' interaction $\dM \times \dM_{ij} \times \dM_{ij}$, as well as the cubic $\dM \times \dM_{ij} \times \dS_{i}$ interaction that also enters at 4PN order. The new feature in the radiative algorithm is that, from the quadratic order onwards, we must apply the gauge transformation defined by Eq.~\eqref{eq:xin}. This gauge transformation is zero for the interactions $\dM \times \dM$ and $\dM \times \dM_{ij}$, hence to control the tails-of-memory we only need the gauge vector for the quadrupole-quadrupole interaction $\dM_{ij} \times \dM_{ij}$. We find~\cite{B98quad} that the source term reads
\begin{align}
	\sigma_{\dM_{ij} \times \dM_{ij}} %&= n_{ijkl} \,M_{ij}^{(3)}M_{kl}^{(3)} - 4 n_{ij} \,M_{ik}^{(3)}M_{jk}^{(3)} + 2 M_{ij}^{(3)}M_{ij}^{(3)} \,,\\
	&= \hat{n}_{ijab} \,M_{ij}^{(3)}M_{ab}^{(3)} - \frac{24}{7} \hat{n}_{ij} \,M_{ia}^{(3)}M_{ja}^{(3)} + \frac{4}{5} M_{ab}^{(3)}M_{ab}^{(3)} \,,
\end{align}
hence the gauge vector~\eqref{eq:xin} upon integration explicitly reads
\begin{subequations}
	\begin{align}
		\xi_{\dM_{ij} \times \dM_{ij}}^0 &= \int_{-\infty}^{u} \dd v \!\int_1^{+\infty} \dd x \biggl\{- \frac{1}{2} \hat{n}_{ijab}\,Q_4(x) M_{ij}^{(3)}M_{ab}^{(3)} + \frac{12}{7} \hat{n}_{ij}\,Q_2(x) M_{ia}^{(3)}M_{ja}^{(3)} \nonumber \\
		&\qquad\qquad\qquad\qquad\quad - \frac{2}{5}\,Q_0(x) M_{ab}^{(3)}M_{ab}^{(3)}\biggr\}\bigl(v-r(x-1), \mathbf{n}\bigr)\,, \\
		%%%%%%%%%%%%%%%%%%%%%%%%%%%%%%%%%%%%%%%%%%%%%%%%%%%%%%%%%%%%%%%%%%%
		\xi_{\dM_{ij} \times \dM_{ij}}^i &= \int_{-\infty}^{u} \dd v \!\int_1^{+\infty} \dd x \biggl\{ - \frac{1}{2} \hat{n}_{iabpq} \,Q_5(x) M_{ab}^{(3)}M_{pq}^{(3)} + \frac{16}{9} \hat{n}_{iab} \,Q_3(x) M_{ak}^{(3)}M_{kb}^{(3)} \nonumber \\
		&\qquad\qquad\qquad\qquad\quad - \frac{2}{9} \hat{n}_{abk} \,Q_3(x) M_{ai}^{(3)}M_{bk}^{(3)} - \frac{22}{35} \hat{n}_{i} \,Q_1(x) M_{ab}^{(3)}M_{ab}^{(3)} \nonumber \\
		&\qquad\qquad\qquad\qquad\quad + \frac{24}{35} \hat{n}_{a} \,Q_1(x) M_{ik}^{(3)}M_{ka}^{(3)} \biggr\}\bigl(v-r(x-1), \mathbf{n}\bigr)\,.
	\end{align}
\end{subequations}
For this interaction, the non-linear correction term vanishes, \textit{i.e.} $\Omega_{\dM_{ij} \times \dM_{ij}}^{\mu\nu}= 0$, and it automatically follows that $\Delta_{\dM_{ij} \times \dM_{ij}}^{\mu}= 0$,  $\phi_{\dM_{ij} \times \dM_{ij}}^{\mu}= 0$, and $\mathcal{H}_{\dM_{ij} \times \dM_{ij}}^{\mu\nu}=0$. This means that there are no corrections to the moments due to this interaction, and since $\zeta_{\dM_{ij} \times \dM_{ij}}^{\mu}= 0$, the the total shift vector reduces to  
\begin{equation}
\varphi_{\dM_{ij} \times \dM_{ij}}^{\mu}=\xi_{\dM_{ij} \times \dM_{ij}}^{\mu}\,.
\end{equation}
When computing the cubic metric, we need also the quadratic $M\times M$ interaction, and this one is trivially computed: the gauge vector vanishes, $\varphi_{\dM \times \dM}^\mu = 0$, while the only non-zero component of the correction term is $\Omega_{\dM\times\dM}^{00}= \frac{4M^2}{r^2}$.

A straightforward dimensional analysis shows that the only cubic interactions which can enter the relation between the canonical moments $\dM_{ij}$ and $\dMbar_{ij}$ up to the 4PN order in the frame of center-of-mass (for which $\dM_{i}=\dMbar_{i}=0$), are precisely the tail-of-tail interaction $\dM \times \dM \times \dM_{ij}$ already computed in the previous section, which is at 3PN order, and the cubic interactions $\dM \times \dM_{ij} \times \dM_{ij}$ and $\dM \times \dM_{ij} \times \dS_{i}$. Hence we limit ourselves to 4PN order. Implementing the calculation using the technical formulae displayed in Appendix~\ref{sec:Retard} we then find that the complete relation up to 4PN order between the moments is (restoring at this point the factors $1/c$)
\begin{equation}\label{eq:deltaMijbarres}
	\begin{aligned}
		\dMbar_{ij} =
		\dM_{ij}
		&\
		+ \left[-\frac{26}{15}+2 \ln\left(\frac{r_0}{b_0}\right)\right]\frac{G\,\dM}{c^3}\,\dM_{ij}^{(1)}\\
		& \ +
		\left[\frac{124}{45}-\frac{52}{15} \ln\left(\frac{r_0}{b_0}\right) +2  \ln^2\left(\frac{r_0}{b_0}\right)  \right]\frac{G^2\dM^2}{c^6}\,\dM_{ij}^{(2)}\\
		& \
		+ \frac{G^2 \dM}{c^8}
		\bigg[-\frac{8}{21}\,\dM^{}_{a\langle i}\dM_{j\rangle a}^{(4)}
		-\frac{8}{7}\,\dM^{(1)}_{a\langle i}\dM_{j\rangle a}^{(3)}
		-\frac{8}{9}\,\dM^{(3)}_{a\langle i}\dS_{j\rangle \vert a}^{}\bigg]+\mathcal{O}\left(\frac{1}{c^9}\right)\,,
	\end{aligned}
\end{equation}
where have posed $\dS_{i\vert j} \equiv \varepsilon_{ijk}\dS_k$ for the angular momentum, and angular brackets denote the STF projection. We can easily invert the previous relation: using the fact that for the conserved (ADM) quantities $\dM=\dMbar$ and $\dS_i = \dSbar_i$, we find 
\begin{equation}\label{eq:deltaMijres}
	\begin{aligned}
		\dM_{ij} =
		\dMbar_{ij}
		&\
		+ \left[\frac{26}{15}-2 \ln\left(\frac{r_0}{b_0}\right)\right]\frac{G\,\dM}{c^3}\,\dMbar_{ij}^{(1)}\\
		& \ +
		\left[\frac{56}{225}-\frac{52}{15} \ln\left(\frac{r_0}{b_0}\right) +2  \ln^2\left(\frac{r_0}{b_0}\right)  \right]\frac{G^2\dM^2}{c^6}\,\dMbar_{ij}^{(2)}\\
		& \
		+ \frac{G^2 \dM}{c^8}
		\bigg[\frac{8}{21}\,\dMbar^{}_{a\langle i}\dMbar_{j\rangle a}^{(4)}
		+\frac{8}{7}\,\dMbar^{(1)}_{a\langle i}\dMbar_{j\rangle a}^{(3)}
		+\frac{8}{9}\,\dMbar^{(3)}_{a\langle i}\dSbar_{j\rangle \vert a}^{}\bigg]+\mathcal{O}\left(\frac{1}{c^9}\right)\,.
	\end{aligned}
\end{equation}
Note that the correction terms we find in~\eqref{eq:deltaMijbarres}--\eqref{eq:deltaMijres} are purely local (no hereditary integrals at this order). Recall that they depend on the two constant scales: $r_0$ the Hadamard regularization scale (or renormalization scale~\cite{GRoss10, GRR12}) entering into both harmonic and radiative constructions of the metric (and supposed to be identical in the two constructions), and $b_0$ the scale used in the harmonic construction when defining observable quantities at infinity, and equivalently entering the gauge transformation between the harmonic and radiative linear metrics, see~\eqref{eq:hrad1}--\eqref{eq:xi1}. 

Finally, with the result~\eqref{eq:deltaMijbarres} in hand, we have been able to reexpress our result for \emph{tails-of-memory} an \emph{spin-quadrupole tails}~\cite{TB23_ToM}, which is performed following the radiative algorithm in terms of the quadrupole moment $\dMbar_{ij}$, in terms of the moment $\dM_{ij}$, therefore matching previous results obtained in harmonic coordinates.

\acknowledgments

We acknowledge discussions with Laura Bernard and Guillaume Faye. F.L. received funding from the European Research Council (ERC) under the European Union’s Horizon 2020 research and innovation programme (grant agreement No 817791). 

\appendix

\section{Formulae for the retarded integrals}
\label{sec:Retard}

During the application of the procedure described in Sec.~\ref{sec:General}, we have to compute some retarded integrals given by Eq.~\eqref{eq:defXY}. We quickly recall the method we follow, see Sec.~IV.B in~Ref.~\cite{MQ4PN_jauge}. Because of the explicit factors $B$ and $B^2$ in their source terms, the retarded integrals~\eqref{eq:defXY} will be non-zero only when they develop a pole (or a double pole) when $B\to 0$. In turn this means that they depend only on the behaviour of the corresponding source in the ``near-zone'', \emph{i.e.}, when $r\to 0$. 

Thus the first task is to expand the source term when $r\to 0$. This is straightforward except when the source contains a hereditary tail integral, say
\begin{equation}\label{eq:defFm}
	\mathcal{F}_m(r,t) = \int_1^{+\infty}\!\!\dd x \,Q_m(x)\,F\left(t-r x/c\right)\,,
\end{equation}
where $F(t)$ denotes a component of the mass quadrupole moment. The formula needed to handle this case has been developed in the App.~A of~\cite{MQ4PN_jauge}: the near-zone expansion of~\eqref{eq:defFm}, valid to any order when $r\to 0$, reads
\begin{subequations}
	\begin{align}\label{eq:resFmfinal}
	\mathcal{F}_m 
	= & \sum_{i=0}^{+\infty} \beta_{i}^m\,\frac{(-)^i}{i!}\left(\frac{r}{c}\right)^i F^{(i)}(t)\\
	& + \sum_{j=0}^{+\infty} \frac{(-)^m\,c_{j}^m}{(m+2j)!}\left(\frac{r}{c}\right)^{m+2j}\int_0^{+\infty}\!\!\dd\tau\biggl[\ln\left(\frac{c\tau}{2r}\right) - H_{m+j} + 2 H_{2m+2j+1} \biggr]\, F^{(m+2j+1)}\left(t-\tau\right)\,,\nonumber
\end{align}
where $H_q$ is the harmonic number, and the coefficients are
\begin{align}
	\!\!\beta_{i}^{m} &= \sum_{k=0}^{m-1} {\genfrac{(}{)}{0pt}{}{i}{k}}\,2^k(k!)^2\frac{(m-k-1)!}{(m+k+1)!}
	+ \sum_{k=m}^{i} {\genfrac{(}{)}{0pt}{}{i}{k}}\,2^k(k!)^2(-)^{m+k}\frac{H_{k-m} - H_{k+m+1}}{(k-m)!(k+m+1)!}\,,\\
	c_{j}^m &= \frac{2^m}{j!}\frac{(m+2j)!(m+j)!}{(2m+2j+1)!}
\end{align}
\end{subequations}
with the binomial symbol $\genfrac{(}{)}{0pt}{}{i}{k}=0$ whenever $i<k$.

Once the near-zone expansion of the source is achieved, it remains to apply the following formulae, most of them being already exposed in the Sec.~IV.B of~\cite{MQ4PN_jauge}, but which we had to generalize in order to include higher powers in the logarithms. Thus the generalization of Eqs. (4.17) in~\cite{MQ4PN_jauge} is
\begin{subequations}\label{eq:solZP}
\begin{align}
	\FPprop\bigg[ \left(\frac{r}{r_0}\right)^B\!B^b \,\frac{\ln^a\! r\,\hat{n}_L}{r^p}\,G\left(t\right)\bigg]
	= e_{p}^{\ell}\,\alpha_{a,b}\,\hat{\partial}_L\bigg[\frac{G^{(p-\ell-3)}\big(t-r/c\big)}{r\,c^{p-\ell-3}}\bigg]\,,
\end{align}
where $e_{p}^{\ell} = 0$ when $p-\ell -3$ is not an even natural integer, and otherwise: 
\begin{align}\label{eq:epell}
	e_{p}^{\ell} = \frac{(-)^p}{(p-\ell-3)!!\,(p+\ell-2)!!}\qquad\text{(when $p=\ell +3 + 2j$ with $j\in \mathbb{N}$)}\,,
\end{align}
\end{subequations}
and where, for the values we need in this paper: $\alpha_{0,1} = 1$, $\alpha_{0,2} = 0$, $\alpha_{1,1} = \ln r_0$, $\alpha_{1,2} = -1$, $\alpha_{2,1} = (\ln r_0)^2$ and $\alpha_{2,2} = -2 \ln r_0$.

It is also very useful to dispose of the similar formula but with a source term which is ``exact'', \emph{i.e.}, not Taylor-expanded in the near zone. In this case, generalizing Eq.~(4.11) of Ref.~\cite{MQ4PN_jauge}:
\begin{subequations}\label{eq:solexact}
\begin{align}
	\FPprop\bigg[ \left(\frac{r}{r_0}\right)^B\!B^b \,\frac{\ln^a\! r\,\hat{n}_L}{r^p}\,G\left(t-r/c\right)\bigg]
	= f_{p}^{\ell}\,\alpha_{a,b}\,\hat{\partial}_L\bigg[\frac{G^{(p-\ell-3)}\big(t-r/c\big)}{r\,c^{p-\ell-3}}\bigg]\,,
\end{align}
where $f_{p}^{\ell} = 0$ if $p-\ell -3 <0$ and otherwise: 
\begin{align}\label{eq:fpell}
	f_{p}^{\ell} = \frac{(-)^p\,2^{p-3}\,(p-3)!}{(p-\ell-3)!\,(p+\ell-2)!}\qquad\text{(when $p\geqslant \ell +3$)}\,,
\end{align}
\end{subequations}
while the values of $\alpha_{a,b}$ remain the same as in~\eqref{eq:solZP}. One can naturally recover the ``exact'' result~\eqref{eq:solexact} from the near-zone result~\eqref{eq:solZP} by performing a Taylor expansion at the level of the source and a subsequent formal resummation.

\bibliography{ListeRef_rad2harm.bib}

\end{document}